\documentclass[a4paper,11pt]{article}
\pdfoutput=1 

\usepackage{jcappub} 

\usepackage[T1]{fontenc} 
\usepackage{xcolor}
\usepackage{subfigure}  

\title{\boldmath Testing Gauss-Bonnet Gravity with DESI BAO Data}

\author[a]{Praveen Kumar Dhankar,}
\author[b]{Dalale Mhamdi,}
\author[c,d,1]{Albert Munyeshyaka,\note{Corresponding author.}}
\author[e]{Darshan Kumar,}
\author[c,f]{Joseph Ntahompagaze,}
\author[b]{Taoufik Ouali}

\affiliation[a]{Symbiosis Institute of Technology, Nagpur Campus, Symbiosis International (Deemed University), Pune-440008, Maharashtra, India}
\affiliation[b]{Laboratory of Physics of Matter and Radiations, Mohammed I University, BP 717, Oujda, Morocco}
\affiliation[c]{Rwanda Astrophysics, Space and Climate Science Research Group, University of Rwanda, College of Science and Technology, Kigali, Rwanda}
\affiliation[d]{Kibogora Polytechnic, Faculty of Education, Western Province, Rwanda}
\affiliation[e]{Institute for Gravitational Wave Astronomy, Henan Academy of Sciences, Zhengzhou-450046, Henan, China}
\affiliation[f]{Department of Physics, College of Science and Technology, University of Rwanda, Kigali, Rwanda}

\emailAdd{pkumar6743@gmail.com}
\emailAdd{dalale.mhamdi@ump.ac.ma}
\emailAdd{munalph@gmail.com}
\emailAdd{kumardarshan@hnas.ac.cn}
\emailAdd{ntahompagazej@gmail.com}
\emailAdd{t.ouali@ump.ac.ma}

\abstract{In the present paper, we observationally constrain \( f(G) \) gravity at the background level using Type Ia supernovae from the Pantheon Plus (PP) sample, cosmic chronometer (CC) data, and the recent Baryon Acoustic Oscillation (BAO) measurements released by DESI. For the analysis, we consider two combinations of datasets: (i) PP + CC, and (ii) PP + CC + DESI BAO. In both cases, we determine the best-fit parameters by numerically solving the modified Friedmann equations for two distinct \( f(G) \) models, namely the power-law and exponential forms. This is achieved through Markov Chain Monte Carlo (MCMC) simulations. To assess the statistical significance of the \( f(G) \) models, we employ both the Akaike Information Criterion (AIC) and the Bayesian Information Criterion (BIC). Our results show that both \( f(G) \) models are statistically favored over the standard \( \Lambda \)CDM model. Notably, the exponential model exhibits an additional future transition at redshift \( z \approx -0.1 \), indicating a possible return to a decelerating phase. This distinctive behavior sets it apart from both the power-law model and the \( \Lambda \)CDM scenario, which predict continued acceleration into the future.}

\begin{document}
\maketitle
\flushbottom
\section{Introduction}\label{intro}
In the past years, a significant number of observations have supported general relativity (GR) as a very successful theory of gravity. This was confirmed by different predictions such as the deflection of light by the Sun's gravitational field \cite{dyson1920ix}, the perihelion motion of Mercury \cite{janssen2021einstein}, the existence of gravitational waves \cite{abbott2016ligo}, and gravitational redshift \cite{landau2013classical} as discussed in \cite{ishak2019testing,hazarika2024f}. However, recent observations revealed different challenges that GR has been facing at both quantum and larger scale levels such as its failure to explain recent cosmic acceleration. \\The observations like type Ia supernovae, large-scale structure and the measurements of the cosmic microwave background anisotropies \cite{riess1998observational,perlmutter1999measurements,tonry2003cosmological,spergel2007three,aghanim2021erratum} provided enough evidence to rule out GR's ability to describe the dynamics of the universe at large scale during its late time evolution. This has led to the proposition of considering alternative theories of gravity or the consideration of different types of fluids with negative pressure such as the scalar field \cite{copeland2006dynamics}, chaplygin gas \cite{saadat2013viscous,sahlu2019chaplygin,gadbail2022generalized,sahlu2023confronting} and the cosmological constant, $\Lambda$ \cite{carroll2001cosmological}. The use of a cosmological constant $\Lambda$ produced the $\Lambda$CDM model consistent with the current cosmological observations. This model assumes a fine-turned cosmological constant that drives the accelerating expansion of the universe \cite{carroll2001cosmological}. One way of dealing with cosmological observations requires the use of geometric probes, such as type Ia supernova \cite{riess1998observational,perlmutter1999measurements,li2023forecast}, the cosmic microwave background (CMB) angular power spectrum \cite{hinshaw2013nine,douspis2019tension}, and the Baryon acoustic oscillations \cite{kazantzidis2021sigma,alam2017clustering}. Although the $\Lambda$CDM model is consistent with recent cosmological observations, it faces different challenges such as the cosmological constant problem and the problems of $H_0$ and $\sigma_8$ tensions. This has led to the consideration of modified theories of gravity as alternative theories to describe the dynamics of the universe at both early inflationary and late-time. \\In recent years, different modified theories of gravity have been explored, such as the $f(R)$, $f(T)$, $f(Q)$, $f(G)$ theories of gravity \cite{nashed2024constraining,escamilla2024f,pawar2024two,gadbail2024modified,maurya2024modified,makarenko2017asymptotic,nojiri2007introduction,munyeshyaka2024covariant,Red2}, to name but a few, where $R$, $T$, $Q$ and $G$ are the Ricci scalar, torsion scalar, nonmetricity scalar, and Gauss-Bonnet invariant, respectively. One of the common features of these theories is that they can unify both the inflationary era of the universe and the late-time era in a similar way to the $\Lambda$CDM. The most challenging issue in each modified theory of gravity is perhaps determining the viable functional form of the model. Although some general analysis such as the existence of Noether symmetries, the absence of ghosts, the stability of perturbations can be extracted through theoretical arguments, there is a need to confront the theoretical model with observations \cite{anagnostopoulos2019bayesian}. In this context, different observational works have been conducted using solar system data \cite{iorio2012solar} and cosmological data \cite{capozziello2015transition,bonici2019constraints,pan2024interacting}. Furthermore, the confrontation with cosmological data used mainly supernovae type Ia data, cosmic microwave background (CMB), baryonic acoustic oscillations (BAO) and Hubble data observations to constrain background evolution \cite{anagnostopoulos2019bayesian}. Ref. \cite{Valentino} addressed observational tensions in cosmology with systematics and fundamental physics and discussed the most promising array of potential new physics that may be observable in upcoming surveys. The authors discussed the growing set of novel data analysis approaches that can go beyond traditional methods to test physical models. In \cite{wang2024constraining}, the authors constrained cosmological physics with DESI BAO observations and suggested that dark energy may be transferred from dark matter to dark energy in the dark sector of the universe. The work presented in \cite{pang2024reevaluating} reevaluated the tension $H_{0}$ with the joint analysis of non-Planck CMB and DESI BAO and showed argued that by combining DESI BAO data + non-Planck CMB measurements, there is a more stringent constraint on the Hubble constant as well as reducing the significance of the Hubble tension. In \cite{zheng2024cosmological}, the authors presented cosmological constraints on dark-energy models using DESI BAO measurements and showed that combining these newly late universe probes can significantly improve constraints on cosmological parameters, as this combination can effectively break parameter degeneracies.
The authors in \cite{sokoliuk2023impact} discussed the impact of $f(Q)$ gravity on the large-scale structure and performed Markov Chain Monte Carlo (MCMC) sampling  on OHD/BAO/Pantheon data sets and constrained parameter space.\\
In addition, the modified theory of $f(Q)$ gravity when limited to observational data leads to interesting cosmological phenomenology at the background level \cite{yang2024data,sahlu2024constraining,mandal2020cosmography,Red}. In the work carried out by \cite{enkhili2024cosmological}, the authors constrained the cosmological dynamical dark energy model in the context of $f(Q)$ gravity. In addition, different authors have been testing this kind of theory against various observational data, which include both background and perturbation observations \cite{sahlu2024constraining,mhamdi2024cosmological,mhamdi2025cosmological,barros2020testing,lazkoz2019observational,anagnostopoulos2021first}, and revealed that $f(Q)$ gravity can challenge the standard $\Lambda$CDM scenario. In \cite{lazkoz2019observational}, the authors constrained the $f(Q)$ gravity using different observational probes and confirmed that the approach used can provide a different perspective on the formulation of observationally reliable alternative model of gravity. The research carried out in the works presented in \cite{mhamdi2024constraints,mhamdi2024cosmological,sahlu2025structure,sahlu2024constraining} treated $f(Q)$ gravity using the mentioned observational data and the constrained model parameters.\\ 
Several studies have investigated modified Gauss-Bonnet gravity $f(G)$ as a promising alternative to general relativity to explain various phases of cosmic evolution. In Ref. \cite{makarenko2017asymptotic}, the authors reconstructed an $f(G)$ model using an exponential scale factor and showed that it can realise a bouncing cosmology, with asymptotic analysis revealing behavior consistent with late-time acceleration. Another work \cite{munyeshyaka2024covariant} examined cosmological perturbations in $f(G)$ gravity with a scalar field using the $1+3$ covariant formalism, demonstrating that the growth of matter over-densities deviates significantly from $\Lambda$CDM predictions, especially due to modifications introduced by the scalar field. Ref. \cite{nojiri2005modified} focused on the theoretical viability of $f(G)$ models, showing that they can satisfy solar system tests and naturally account for both the transition from deceleration to acceleration. In addition, Ref. \cite{lee2020viable}  explored the reconstruction of viable $f(G)$ models from observational data, emphasising their consistency with cosmological evolution. Ref. \cite{Ghost} developped a framework for the Gauss-Bonnet gravity. This involves using Lagrange multipliers technique and by means of constraints. The authors then explored the modifications to the Newtonian law of gravity by means of the ghost-free $f(G)$ theory. In Ref. \cite{Dark}, the authors demonstrated that the $f(G)$ is a viable theory compliant with the solar system constraints and addressed the issue of a possible solution of the hierarchical problem in modified theories of gravity.
 Collectively, these works illustrate the potential of $f(G)$ gravity to describe both the background  and perturbative aspects of the universe beyond general relativity.\\

Motivated by the mentioned works, we will constrain model parameters in the context of modified Gauss-Bonnet gravity. For pedagogical purpose, we incorporate two different $f(G)$ gravity models, which include the power-law $f(G)$ model given by $f(G) = \alpha G^\beta$, dubbed (model I) and an exponential $f(G)$ model given by $f(G) = \alpha G_0(1 - e^{-p G/G_0})$, dubbed (model II) functions of Gauss-Bonnet invariant, $G$, where  $\alpha$, $\beta$, $G_{0}$, and $p$ are constants. After obtaining the modified Friedmann equation, we aim to constrain parameters of the defined $f(G)$ gravity models by means of Bayesian analysis of Hubble measurement, Pantheon plus supernova type Ia and the DESI BAO measurement. In order to achieve this aim, we numerically solve the modified Friedmann equations under the assumption that the Universe is exclusively composed of dust matter, where the equation of state parameter vanishes.\\

The next task is to use MCMC analysis to constrain model parameters resulting from the use of $f(G)$ gravity models. In so doing, we compute the corner (triangular) plots and present the mean value parameters corresponding to each data sets namely: (i) Pantheon plus supernova measurements (ii) Hubble parameter measurements and iii) and DESI BAO data. For further analysis, we use joint analysis of iv) Pantheon plus and cosmic chronometers (PP+CC) and v)  Dataset consists of Pantheon plus, cosmic chronometers, and DESI BAO data (PP+CC+DESI BAO)  in the context of $f(G)$ gravity theory for each model. The last part of this work considered statistical analysis where we obtain statistical results for two $f(G)$ models, which helps to compare the models with $\Lambda$CDM. In order to achieve this task, we use the Akaike information criteria (AIC) and the Bayesian information criteria (BIC). After getting the best fit parameters  and statistically compare the models with $\Lambda$CDM, we examine how these $f(G)$ gravity models enhance the current cosmic acceleration by employing the deceleration parameter $q$ within the framework of $f(G)$ gravity.\\

The rest of this paper is organized as follows: In Section~\ref{ba}, we present the mathematical framework, where the cosmological equations are discussed in the context of \( f(G) \) gravity. In Section~\ref{da}, we describe the data and methodology employed to constrain the two \( f(G) \) models. Section~\ref{re} presents and discusses the results obtained, while Section~\ref{con} is reserved for the conclusions.

\section{The $f(G)$ cosmology }\label{ba}
In the present section, we describe the modified Gauss-Bonnet gravity, where the Friedmann equation will be modified with respect to the Gauss-Bonnet gravity. After obtaining such modified Friedmann equation, we will define specific $f(G)$ models and present normalised Friedmann equation for each model.
\subsection{Background evolution in the context of modified Gauss-Bonnet gravity}
The gravitational action involving normal matter assisted by $f(G)$ gravity is  presented as \cite{nojiri2011unified,nojiri2005modified,li2007cosmology,venikoudis2022late}
\begin{eqnarray}
 S=\int d^{4}x\sqrt{-g}\Big(\frac{R}{2\kappa^{2}}+\frac{f(G)}{2}+\mathcal{L}_{m}\Big)\;
 \label{eq1},
\end{eqnarray}
where $R$ is the Ricci scalar, $f(G)$ represents an arbitrary function of the Gauss-Bonnet invariant $G$, $\mathcal{L}_{m}$ is the usual matter Lagrangian, $g$ is the determinant of the metric $g^{\mu \nu}$ and $\kappa$ is the gravitational constant. 
For the case $f(G)=G$, $\int d^{4}x\sqrt{-g}G=0$, we recover the  gravitational action for GR. In this context, Eq. (\ref{eq1}) produces gravitational field equations represented as
\begin{eqnarray}
 G_{\mu \nu} \equiv R_{\mu \nu}-\frac{1}{2}g_{\mu \nu}R=8\pi G_{N}T^{tot}_{\mu \nu},\label{eq2}
\end{eqnarray}
where $T^{tot}_{\mu \nu}$ represents the energy momentum tensor of the total fluids (matter and Gauss-Bonnet fluids). For a perfect fluid, the energy-momentum tensor is given by
\begin{eqnarray}
 T^{tot}_{\mu \nu}=(\rho+p)u_{\mu}u_{\nu}+pg_{\mu \nu},
\end{eqnarray}
where, $\rho$ and $p$ are the energy density and isotropic pressure, respectively and $u^{\mu}$ is the $4$-velocity. The adopted spacetime signature is $(-,+,+,+)$ and unless stated otherwise, we
use $\mu$, $\nu$ . . . $= 0$, $1$, $2$, $3$, and $8\pi G_{ N} = c = 1$, where $G_{ N}$ is the gravitational constant
and $c$ is the speed of light. 
For a flat, homogeneous, and isotropic Friedmann-Robertson-Walker (FRW) universe, the metric is presented as 
\begin{eqnarray}
ds^{2}=-dt^{2}-a^{2}\Big(dx^{2}+dy^{2}+dz^{2}\Big),
\end{eqnarray} where $a(t)$ is the cosmological scale factor. The Ricci scalar and Gauss-Bonnet invariant are presented as $R=6\Big(\dot{H}+2H^{2}\Big)$ and $G=24H^{2}\Big(\dot{H}+H^{2}\Big)$, respectively. The corresponding modified Friedmann and Raychaudhuri equations become
\begin{eqnarray}
&& 3H^{2}=\rho_{m}+\rho_{r}+\rho_{G},\label{eq2.5}\\
&&\Big(2\dot{H}+3H^{2}\Big)=-p_{m}-p_{r}-p_{G},
\end{eqnarray} respectively, where
\begin{eqnarray}
&&\rho_{G}=-\frac{1}{2}f+\frac{1}{2}Gf',\label{eq2.7}\\
&&p_{G}=-\frac{1}{2}Gf'+\frac{1}{2}f-12H\dot{f}'H^{2}.
\end{eqnarray} 
The energy density for baryonic matter and radiation are given by $\rho_{m}=\rho_{m0}(1+z)^{3}$, $\rho_{r}=\rho_{r0}(1+z)^{4}$ and $\rho_{G}$ is the energy density resulting from the Gauss-Bonnet fluid. The $p_{m}$ and $p_{G}$ are the respective pressures. The  $f'$ is the usual partial derivative with respect to the Gauss-Bonnet invariant. The pressure is related to the energy density by the equation of state parameter $w$ as $p=w\rho$,  where $p$ and $\rho$ are the pressure and energy density, respectively. For $w=0$, we have pressure-less matter (dust) in the universe, where the matter density is low. For $w=\frac{1}{3}$, the linear relationship between pressure and energy density describes the radiation era in the early universe, characterised by a high density. The continuity equation can be represented individually as
\begin{eqnarray}
&& \dot{\rho}_{m}+\theta \Big(\rho_{m}+p_{m}\Big)=0,\\
&&\dot{\rho_{G}}+\theta \Big(\rho_{G}+p_{G}\Big)=0,
\end{eqnarray}
where $\theta=3H$, $H$ is the Hubble parameter. Define the dimensionless Hubble parameter $E$ as $E^{2}=\frac{H^{2}}{H^{2}_{0}}$, and use Eqs. (\ref{eq2.5}) and  (\ref{eq2.7}), we get
\begin{eqnarray}
E^{2}\equiv\frac{H^{2}}{H^{2}_{0}}=\Omega_{m0}\Big(1+z\Big)^{3}+\Omega_{r0}\Big(1+z\Big)^{4}+\Omega_{f0}y(z,r)\label{eq2.11}
\end{eqnarray}
where $H_{0}$, $\Omega_{r0}=\frac{\rho_{r0}}{3H^{2}_{0}}$, $\Omega_{m0}=\frac{\rho_{m0}}{3H^{2}_{0}}$ and $\Omega_{f0}=\frac{\rho_{f0}}{3H^{2}_{0}}$ are respectively, the present Hubble parameter, the normalised energy density of radiation, matter and Gauss-Bonnet. We have also assumed that currently $y(z=0,r=1)=1$, and $r$ is the model parameter. Hence $\Big(1-\Omega_{m0}-\Omega_{r0}\Big)=\Omega_{f0}$. It can be shown that from Eq. (\ref{eq2.5}) and  (\ref{eq2.7})
\begin{eqnarray}
y(z,r)=\frac{1}{6H^{2}_{0}\Omega_{f0}}\Big[Gf'-f\Big]\label{eq2.12}
\end{eqnarray}
The values of $y(z,r)$ will be determined from specific form of $f(G)$ models considered in the next subsection. 
\subsection{Specific $f(G)$ models}
In this part, we  explore two different cosmological $f(G)$ models under $f(G)$ gravity theory. We consider  two specific forms of $f(G)$ gravity namely power law and exponential $f(G)$ models thereafter analyse its cosmological implications on the background evolution using different observational data sets.
\subsubsection{Power-law $f(G)$ model}
Let us consider a generic power law $f(G)$ gravity model (hereafter model I) for pedagogical purpose given by
\begin{eqnarray}
f(G)=\alpha G^{\beta}\label{eq2.13},
\end{eqnarray} where $\alpha$ and $\beta$ are constants. 
For the case $\alpha=1$ and $\beta=1$, the $\Lambda$CDM case is retained. This model resembles the one considered in \cite{nojiri2005modified,lee2020viable}, for the case $a_{1}=\alpha$ and $b_{1}=0$.
From Eq. (\ref{eq2.5}) and (\ref{eq2.7}), the parameter $\alpha$ can be given by $\alpha=\frac{6H^{2}_{0}\Omega_{f0}}{(\beta-1)G_{0}^{\beta}}$, where $G_{0}=24H^{4}_{0}$ and $\Omega_{f0}=1-\Omega_{m}-\Omega_{r}$. Using Eq. (\ref{eq2.5}), (\ref{eq2.7}) and $\alpha$, Eq. (\ref{eq2.12}) for this particular $f(G)$ model becomes
\begin{eqnarray}
y(z,\beta)= E^{4\beta}. 
\end{eqnarray}
Consequently, Eq. (\ref{eq2.11}) can be rewritten as 
\begin{equation}
E^2- \Omega_{fo}E^{4\beta}= \Omega_{m}(1+z)^{3}+\Omega_{r}(1+z)^{4}.
\label{F11}
\end{equation}

 Eq. (\ref{F11}) represents the normalised Friedmann equation in the context of $f(G)$ gravity for the power law model. Let us find the normalised Hubble parameter for the exponential model as below.
\subsubsection{Exponential $f(G)$ model}
 Motivated by exponential $f(R)$ gravity \cite{linder2009exponential}  and $f(T)$ gravity \cite{nesseris2013viable} models, we can construct the exponential $f(G)$ gravity model (hereafter model II) as
 
 \begin{eqnarray}
f(G)=\alpha G_{0}\Big(1-e^{-p\frac{G}{G_{0}}}\Big) \label{eq2.16},
\end{eqnarray} where $\alpha$, $G_{0}$ and $p$ are constants.  Following the same procedures as above, it is straightforward to show that from Eqs . (\ref{eq2.5}) and  (\ref{eq2.7}) 
\begin{equation*}
\alpha=\frac{\Omega_{f0}}{4H^{2}_{0}\Big[e^{-p}\Big(p+1\Big)-1\Big]}
.
\end{equation*}
  From Eq. (\ref{eq2.12}) and using $\alpha$, we get 
\begin{equation}
  y(z,p)=  \frac{e^{-p E^{4}}\Big(pE^{4}+1\Big)-1}{e^{-p}\Big(1+p\Big)-1},
\end{equation} 
which yields
\begin{equation}
E^2- \Omega_{f0} \frac{e^{-p E^{4}}\Big(pE^{4}+1\Big)-1}{e^{-p}\Big(1+p\Big)-1}=   \Omega_{m}(1+z)^{3}+\Omega_{r}(1+z)^{4}.\label{F22}
\end{equation} 
Eq. (\ref{F22}) represents the normalised Friedmann equation for the exponential model. \footnote{In this paper, we are interested in the current cosmic acceleration of the universe. Hence to a good approximation, $\frac{\dot{H}}{H^{2}}\ll 1$ and  we can set $G=24 H^{4}$} This equation and Eq. (\ref{F11}) are crucial in analysing background dynamics by comparing with the observational data sets. In the last part of this work, we will also consider the evolution of Hubble parameter and deceleration parameter ($q$) defined as $q=\frac{\ddot{a}a}{\dot{a}^{2}}$, which can be represented in redshift-space as 
\begin{equation}
q(z) = -1 + (1 + z) \frac{E'(z)}{E(z)},\label{F33}
\end{equation} 
by solving numerically this equation (Eq. (\ref{F33})) using Eqs. (\ref{F11}) and  (\ref{F22}), we can detect the acceleration/deceleration phase for each model.
In this work, we are motivated in the viability of $f(G)$ gravity models in the sense that these defined $f(G)$ models can i) describe the matter and dark energy eras ii) they are consistent with observational data iii) pass the solar system tests and iv) they have stable perturbations.  Although these important studies have not yet been performed for all defined $f(G)$ models, failure of a particular model to pass one of these tests is enough to be ruled out. As shown above, for all these two $f(G)$ models, the model parameters measure the smooth deviation from $\Lambda$CDM. In the next section, we focus on them and apply them to the observational data for model parameters estimation.  Let us first begin our discussion by describing the observational datasets and statistical methods used to compare the models to cosmological observation.
\section{Data and methodology}\label{da}
In this section, we place observational constraints on the two \( f(G) \) gravity models, referred to as Model~I and Model~II. To this end, a detailed statistical analysis is carried out by comparing theoretical predictions of the \( f(G) \) gravity models with cosmological observations. Specifically, the parameters of the two models, \( (\Omega_m, \beta, h) \) for Model~I, and \( (\Omega_m, p, h) \) for Model~II, are constrained using cosmological observations via the Markov Chain Monte Carlo method \cite{mcmc1}. It should be noted that the parameter \( h \) is related to the current Hubble rate by \( h = H_0 / 100 \, \mathrm{km\,s^{-1}\,Mpc^{-1}} \). The observational data sets employed to constrain the \( f(G) \) gravity models include the Pantheon plus \cite{pp}, Hubble parameter measurements \cite{Hm}, and DESI BAO data \cite{DESI}. We consider two combinations of observational data: 
\begin{itemize}
    \item \textbf{Dataset~I}: PP+CC.
    \item \textbf{Dataset~II}: PP+CC+DESI BAO.
\end{itemize}
\subsection{Statistical analysis}
In recent years, cosmological models have been tested using different types of observational data. Comparing these models with observations is the only way to see which ones are supported by evidence. To do this properly, we need to use detailed statistical methods. Since cosmology often relies on Bayesian methods, parameter estimation is usually done using Bayesian inference. In the case of Gaussian errors, the chi-square function, $\chi^2$, and the likelihood function, $\mathcal{L}$, are related according to the relation
\begin{equation}
    \chi^2(\theta) = -2 \ln \mathcal{L}(\theta). 
\end{equation}
we employ both the corrected Akaike Information Criterion (AIC$_c$) \cite{AIC} and the Bayesian Information Criterion (BIC) \cite{BIC}, which are defined respectively as
\begin{equation}
    \mathrm{AIC}_c = \chi^2_{\min} + 2\mathcal{K}_f + \frac{2\mathcal{K}_f(\mathcal{K}_f + 1)}{\mathcal{N}_t - \mathcal{K}_f - 1},
\end{equation} and
\begin{equation}
    \mathrm{BIC} = \chi^2_{\min} + \mathcal{K}_f \ln(\mathcal{N}_t),
\end{equation}
Here, \(\chi^2_{\min}\) denotes the minimum chi-square value, \(\mathcal{K}_f\) represents the number of free parameters, and \(\mathcal{N}_t\) is the total number of data points. The model with the lowest AIC\(_c\) and BIC values is regarded as the most supported by the data and is chosen as the reference model. To quantify the performance of other models relative to the reference, we compute the differences.
\begin{equation}
    \Delta \mathrm{AIC}_c = \mathrm{AIC}_{c,\text{model}} - \mathrm{AIC}_{c,\text{ref}},
\end{equation}
\begin{equation}
    \Delta \mathrm{BIC} = \mathrm{BIC}_{\text{model}} - \mathrm{BIC}_{\text{ref}}.
\end{equation}
The interpretation of $\Delta \mathrm{AIC}_c$ (and analogously $\Delta \mathrm{BIC}$) is as follows: $0 < \Delta \mathrm{AIC}_c < 2$ both models have similar support from the data. $2 < \Delta \mathrm{AIC}_c < 4$: the model with the higher AIC$_c$ is less favored. $4 < \Delta \mathrm{AIC}_c < 6$: there is positive evidence against the model with the higher AIC$_c$. $6 < \Delta \mathrm{AIC}_c < 10$: strong evidence exists against the model with the higher AIC$_c$. $\Delta \mathrm{AIC}_c > 10$: the model with the higher AIC$_c$ is strongly disfavored by the data. A similar interpretation holds for $\Delta \mathrm{BIC}$
\subsection{Datasets}
\begin{itemize}
    \item \textbf{Pantheon Plus} \\
    A recently updated compilation of type Ia supernova data, known as Pantheon plus, has been released~\cite{pp}. This dataset includes a total of 1701 data points derived from 1550 SNe Ia, covering a redshift range of \(0.001 \leq z \leq 2.3\). The chi-square statistic used for fitting cosmological models to the Pantheon plus data is given by
    \begin{align}
        \chi^2_{\text{PP}} = \vec{F}^{\,T} \cdot C^{-1}_{\text{PP}} \cdot \vec{F},
        \label{chi}
    \end{align}
    where \(\vec{F}\) is the vector of the difference between the observed apparent magnitudes \(m_{\text{B}i}\) and the predicted magnitudes from the cosmological model. The covariance matrix \(C_{\text{PP}}\) accounts for both statistical and systematic uncertainties. The theoretical prediction for the distance modulus is given by
    \begin{align}
        \mu_{\text{model}}(z_i) = 5 \log_{10} D_L(z_i) + 25,
    \end{align}
    where \(D_L(z_i)\) is the luminosity distance expressed as
    \begin{align}
        D_L(z_i) = (1 + z_i) \int_0^{z_i} \frac{c}{H(z')} \, dz',
    \end{align}
    with \(c\) denoting the speed of light. A key improvement of the Pantheon plus dataset over its predecessor is the decoupling of the absolute magnitude \(M\) of SN Ia from the Hubble constant \(H_0\). This is achieved by redefining the vector \(\vec{F}\) using distance moduli for SNe Ia in Cepheid-host galaxies, which are independently calibrated by the SH0ES collaboration~\cite{SH0ES}. The modified residual vector \(\vec{F}_i\) is thus defined as
    \begin{align}
        \vec{F}_i = 
        \begin{cases} 
            m_{\text{B}i} - M - \mu^\text{Ceph}_i, & \text{if } i \in \text{Cepheid hosts}, \\
            m_{\text{B}i} - M - \mu_{\text{model}}(z_i), & \text{otherwise},
        \end{cases}
    \end{align}
    where \(\mu^\text{Ceph}_i\) is the independently measured distance modulus of the \(i\)th SNe Ia's Cepheid host, and \(\mu_{\text{model}}(z_i)\) is the model-predicted distance modulus at redshift \(z_i\). Here, \(M\) denotes the absolute magnitude of SNe Ia.
\end{itemize}

\begin{itemize}
    \item\textbf{Cosmic chrnometer:} The Hubble parameter can be expressed as \( H(z) = -\frac{dz}{dt}(1 + z)^{-1} \), providing a means to determine its value directly from observational data. By obtaining the redshift variation \(dz\) through spectroscopic surveys and measuring the corresponding time interval \(dt\), one can compute \(H(z)\) in a model-independent manner. The chi-square function associated with the Hubble parameter measurements is defined as
\begin{equation}
\chi_{\text{CC}}^2  = \sum_{i=1}^{36} \left[\frac{H_{obs}(z_i)-H_{th}(z_i)}{\sigma(z_i)}\right]^2,
\end{equation}
Here, \(H_{\text{obs}}(z_i)\) and \(H_{\text{th}}(z_i)\) denote the observed and theoretical values of the Hubble parameter, respectively, while \(\sigma(z_i)\) represents the observational uncertainty associated with \(H_{\text{obs}}(z_i)\). The cosmic chronometer data, which consists of a total of 36 measurements, are obtained by using the dataset provided in~\cite{mhamdi2024cosmological}.
\end{itemize}
\begin{itemize}
    \item \textbf{DESI BAO:} The recent DESI dataset is constructed from observations of various tracers, including bright galaxy samples (BGS), luminous red galaxies (LRGs), emission line galaxies (ELGs), quasars, and the Ly$\alpha$ forest, covering the redshift range \(0.1 < z < 4.2\)  \cite{DESI}. In this work, we utilize their measurements of the comoving distances $D_M(z)/r_d$ and $D_H(z)/r_d$, where

\begin{equation}
    D_M(z) \equiv \int_0^z \frac{c\, dz'}{H(z')}, \quad D_H(z) \equiv \frac{c}{H(z)}.
\end{equation}
\begin{table}[t!]
\centering
\begin{tabular}{|c|c|c|c|}
\hline
\textbf{Type} & \textbf{Tracer} & \textbf{Redshift} & \textbf{Measurement} \\
\hline
$D_{\mathrm{V}} / r_{\mathrm{d}}$   & BGS             & 0.295 & 7.93  \\
$D_{\mathrm{M}} / r_{\mathrm{d}}$   & LRG1            & 0.51  & 13.62 \\
$D_{\mathrm{H}} / r_{\mathrm{d}}$   & LRG1            & 0.51  & 20.98 \\
$D_{\mathrm{M}} / r_{\mathrm{d}}$   & LRG2            & 0.71  & 16.85 \\
$D_{\mathrm{H}} / r_{\mathrm{d}}$   & LRG2            & 0.71  & 20.08 \\
$D_{\mathrm{M}} / r_{\mathrm{d}}$   & LRG3+ELG1       & 0.93  & 21.71 \\
$D_{\mathrm{H}} / r_{\mathrm{d}}$   & LRG3+ELG1       & 0.93  & 17.88 \\
$D_{\mathrm{M}} / r_{\mathrm{d}}$   & ELG2            & 1.32  & 27.79 \\
$D_{\mathrm{H}} / r_{\mathrm{d}}$   & ELG2            & 1.32  & 13.82 \\
$D_{\mathrm{V}} / r_{\mathrm{d}}$   & QSO             & 1.49  & 26.07 \\
$D_{\mathrm{M}} / r_{\mathrm{d}}$   & Ly$\alpha$ QSO  & 2.33  & 39.71 \\
$D_{\mathrm{H}} / r_{\mathrm{d}}$   & Ly$\alpha$ QSO  & 2.33  & 8.52  \\
\hline
\end{tabular}
\caption{DESI Year-1 data used in this work. The table includes the type of measurement, the tracer type, the redshift and measurement \cite{DESI}.}
\label{tab:DESIdata}
\end{table}
\begin{table}[t!]
\centering
\renewcommand{\arraystretch}{1.2}
\begin{tabular}{lccc}
\hline
\textbf{Parameter} & \textbf{$\Lambda$CDM} & \textbf{$f(G)$ Model I} & \textbf{$f(G)$ Model II} \\
\hline
\multicolumn{4}{c}{\textbf{PP + CC}} \\
$\Omega_{m}$    & $0.282 \pm 0.0129$           & $0.190^{+0.041}_{-0.036}$ & $0.262 \pm 0.014$ \\
$h$           & $0.7141 \pm 0.0087$          & $0.7149 \pm 0.0086$        & $0.7110 \pm 0.0087$ \\
$\beta$       & --                           & $0.260^{+0.079}_{-0.069}$ & -- \\
$p$           & --                           & --                         & $3.97^{+0.38}_{-0.47}$ \\
$M$           & $-19.323 \pm 0.0243$         & $-19.309 \pm 0.0240$       & $-19.307 \pm 0.0240$ \\
\hline
$\chi_{min}^2$      & 1564.13                      & 1553.27                    & 1555.67 \\
\hline
\multicolumn{4}{c}{\textbf{PP + CC + DESI BAO}} \\
$\Omega_{m}$    & $0.282 \pm 0.011$            & $0.230^{+0.027}_{-0.024}$  & $0.275 \pm 0.011$ \\
$h$           & $0.7139 \pm 0.0082$          & $0.7115 \pm 0.0083$        & $0.7085 \pm 0.0084$ \\
$\beta$       & --                           & $0.192^{+0.068}_{-0.062}$ & -- \\
$p$           & --                           & --                         & $4.18^{+0.35}_{-0.40}$ \\
$r_d$         & $143.8 \pm 1.6$              & $142.9 \pm 1.7$            & $143.4 \pm 1.6$ \\
$M$           & $-19.325 \pm 0.023$          & $-19.320 \pm 0.024$        & $-19.317 \pm 0.023$ \\
\hline
$\chi_{min}^2$     & 1578.04                      & 1569.67                    & 1571.31 \\
\hline
\end{tabular}
\caption{Mean values of cosmological parameters for the $\Lambda$CDM model and two $f(G)$ gravity models, obtained using the Pantheon plus and Cosmic Chronometers datasets (PP + CC), and the combined dataset including DESI BAO measurements (PP + CC + DESI BAO).}
\label{Mean}
\end{table}
Additionally, we include the angle-averaged distance measure \(D_V(z)/r_d\), where \(r_d\) denotes the sound horizon at the drag epoch, defined as
\begin{equation}
    D_V(z) \equiv \left[ z D_M(z)^2 D_H(z) \right]^{1/3}.
\end{equation}
The BAO distance measurements from DESI Year-1 data are summarized in Table~\ref{tab:DESIdata}. This dataset includes 12 data points, and five of them are correlated (at redshifts $z = 0.51$, $0.71$, $0.93$, $1.32$, and $2.33$). The corresponding covariance matrix that captures these correlations is provided in \cite{DESI2}.
\end{itemize}
\begin{figure}[h!]
    \centering
    \includegraphics[width=0.9\textwidth]{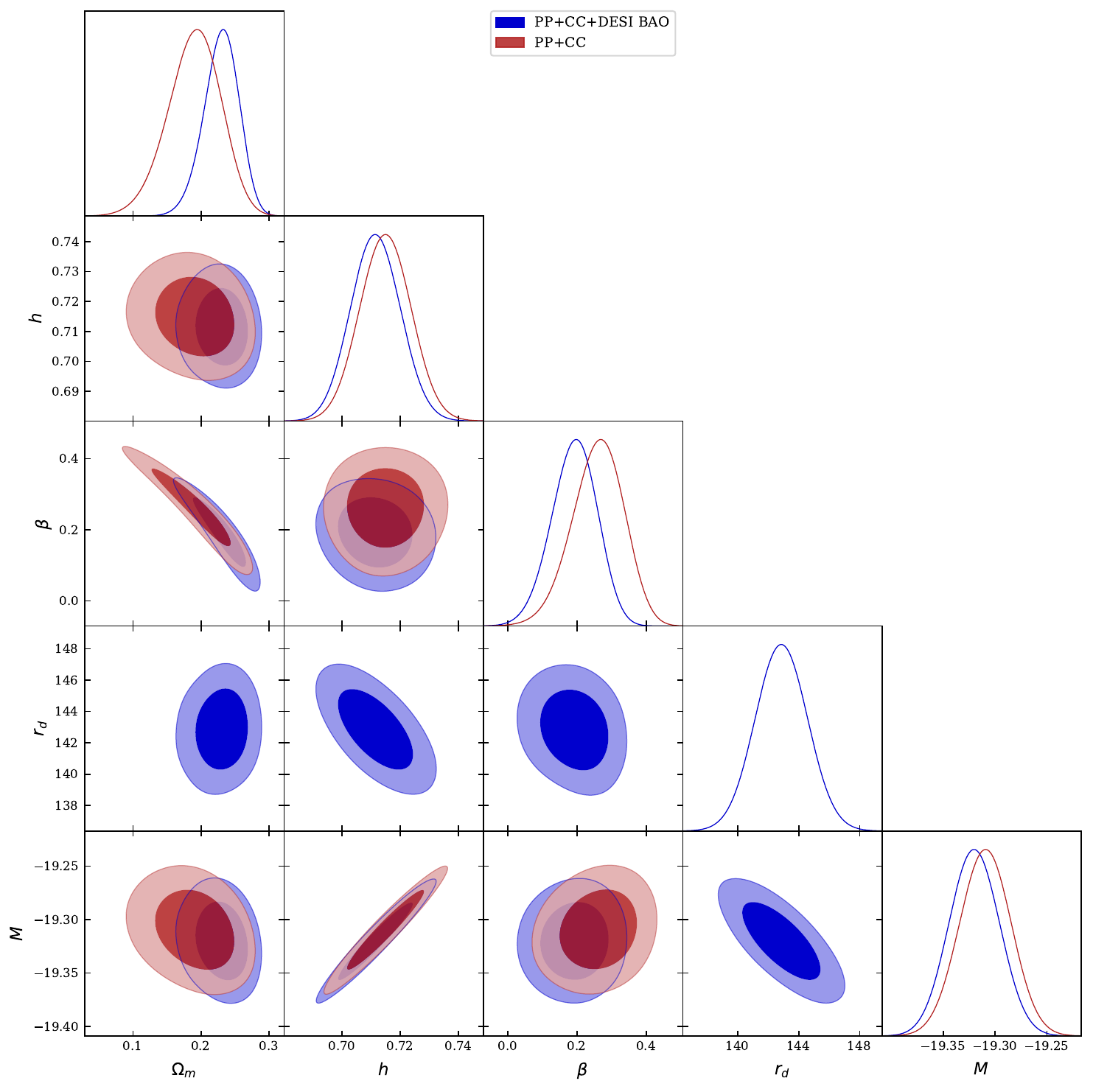}
    \caption{The \(1\sigma\) and \(2\sigma\) confidence contours, along with the posterior distributions, are shown for Model I using two dataset combinations: Dataset I (red) and Dataset II (blue).}
    \label{ca}
\end{figure}
 \begin{figure}[t!]
    \centering
\includegraphics[width=0.9\textwidth]{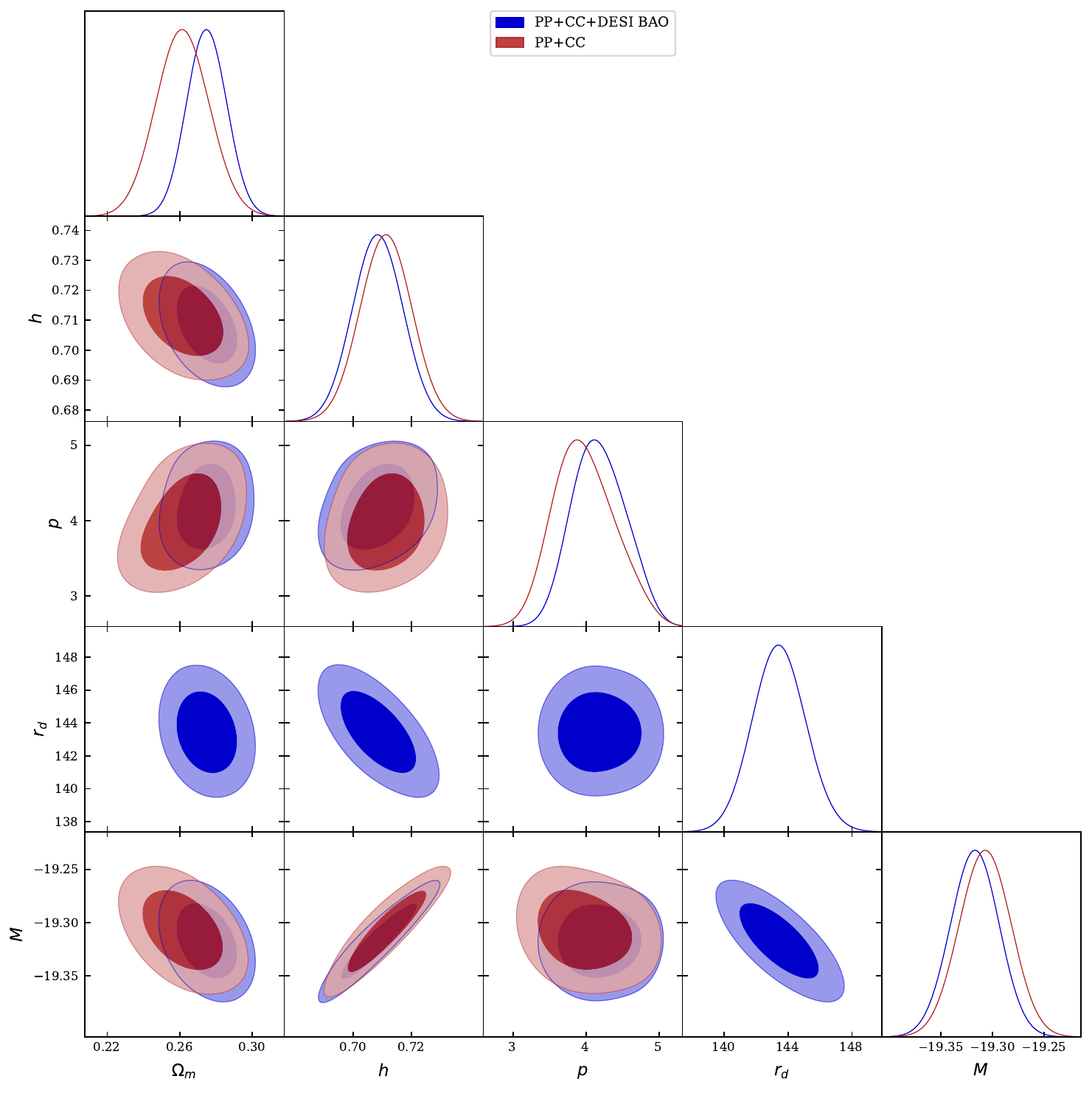}
    \caption{The \(1\sigma\) and \(2\sigma\) confidence contours, along with the posterior distributions, are shown for Model II using two dataset combinations: Dataset I (red) and Dataset II (blue).}
    \label{c2}
\end{figure}
\section{Results and discussions}\label{re}
In this section, we present the results of our analysis of two cosmological models within the $f(G)$ gravity framework. Specifically, we examine Model I and Model II, governed by the Friedmann equations (\ref{F11}) and (\ref{F22}), respectively. Due to the absence of analytical solutions for these equations, we solve them numerically. The free parameters of both models are constrained using Pantheon plus, Cosmic Chronometers, and DESI BAO data. Table. (\ref{Mean}) summarizes the MCMC results for the $\Lambda$CDM model, Model I, and Model II. It displays the mean values of the parameters and their associated $1\sigma$ uncertainties (68\% confidence level), based on two dataset combinations: PP+CC and PP+CC+DESI BAO. The free parameter vector is given by $(\Omega_{m}, h, r_d, M)$ for the $\Lambda$CDM model; $(\Omega_{m}, \beta, h, r_d, M)$ for Model I; and $(\Omega_{m}, p, h, r_d, M)$ for Model II.\\

By constraining the $\Lambda$CDM and $f(G)$ models using the PP+CC+DESI datasets, we find $\Omega_{m} = 0.280 \pm 0.011$ for the $\Lambda$CDM model. This value is higher than those obtained for the $f(G)$ models: $\Omega_{m} = 0.230^{+0.027}_{-0.024}$ for Model I and $\Omega_{m} = 0.275 \pm 0.011$ for Model II, corresponding to deviations of $1.8\sigma$ and $0.32\sigma$, respectively. The Hubble parameter is found to be $h = 0.7139 \pm 0.0082$ km/s/Mpc for $\Lambda$CDM, slightly higher than the values for Model I ($h = 0.7115 \pm 0.0083$ km/s/Mpc) and Model II ($h = 0.7085 \pm 0.0084$ km/s/Mpc). The sound horizon at the drag epoch is $r_d = 143.8 \pm 1.6$ Mpc for $\Lambda$CDM, compared to $r_d = 142.9 \pm 1.7$ Mpc for Model I and $r_d = 143.0 \pm 0.023$ Mpc for Model II. Additionally, the $f(G)$ model parameters are constrained as follows: $\beta = 0.192^{+0.068}_{-0.024}$ for Model I and $p = 4.18^{+0.35}_{-0.40}$ for Model II.\\

\begin{figure}[t!]
    \centering
    \includegraphics[width=0.9\textwidth]{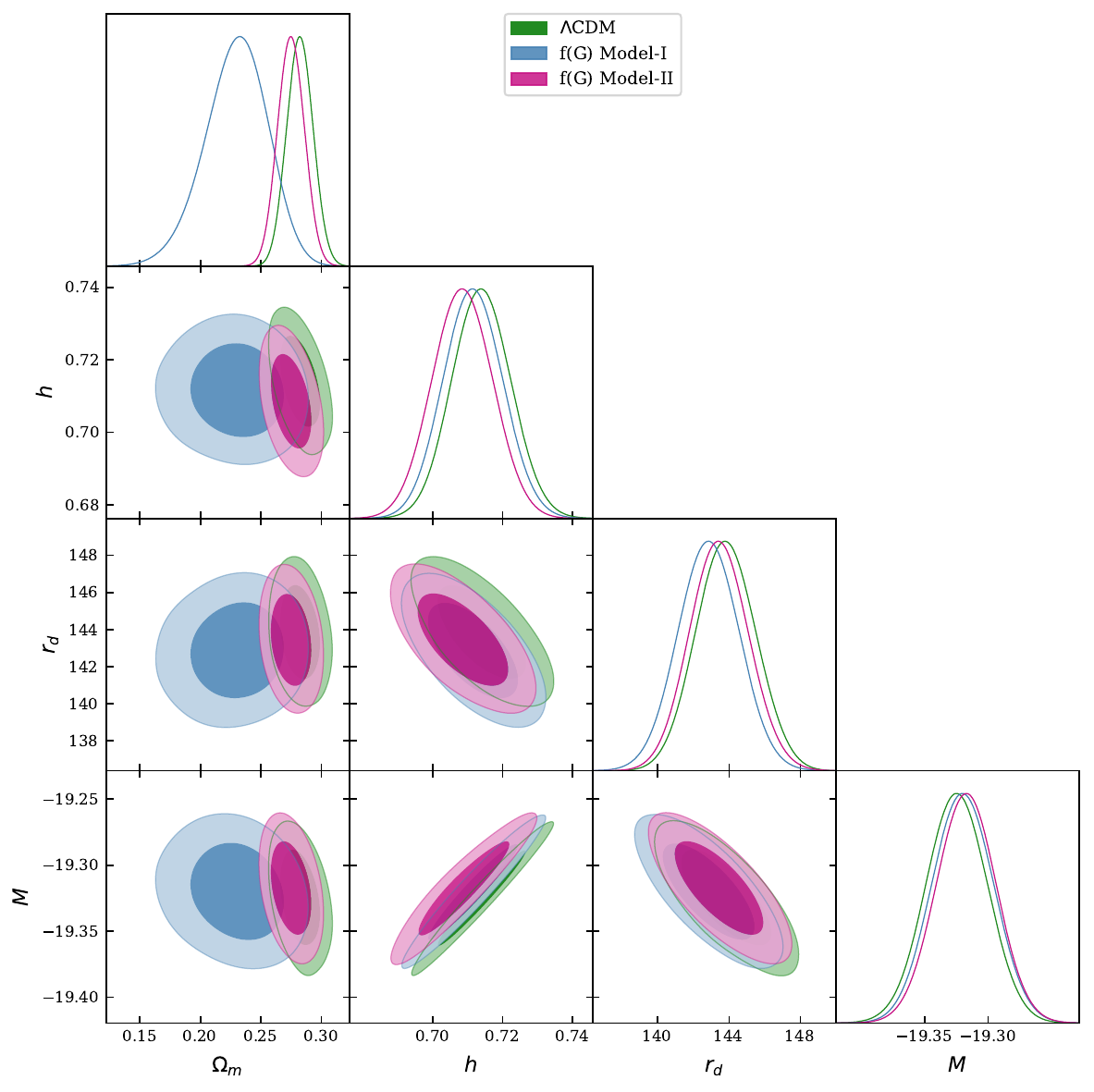}
    \caption{The 1$\sigma$ and 2$\sigma$ confidence contours and the posterior distributions obtained for the Model I, Model II, and $\Lambda$CDM, using Dataset II.}
    \label{c3}
\end{figure}

Using datasets~I and~II, we present the one-dimensional (1D) and two-dimensional (2D) posterior distributions, corresponding to the 68.3\% ($1\sigma$) and 95.4\% ($2\sigma$) confidence levels, in Figs.~\ref{ca} and~\ref{c2} for Model~I and Model~II, respectively. For Model~I, Fig. (\ref{c2}) shows positive correlations in the $(h, M)$ and $(\beta, M)$ planes, and negative correlations in the $(\Omega_{m}, \beta)$, $(r_d, M)$, and $(h, r_d)$ planes. In the case of Model~II, Fig. (\ref{c2}) illustrates positive correlations in the $(h, M)$ and $(\Omega_{m}, p)$ planes, while negative correlations appear in the $(\Omega_{m}, M)$, $(M, r_d)$, $(h, r_d)$, and $(h, \Omega_{m})$ planes. Fig. (\ref{c3}) presents the 1D and 2D posterior distributions for all three models: $\Lambda$CDM (in green), Model~I (in blue), and Model~II (in pink), obtained using Dataset~II. \\

As the three models $\Lambda$CDM, Model~I, and Model~II have different numbers of free parameters, the minimum chi-square value, $\chi^2_{\min}$, alone is insufficient for model comparison. Therefore, we compute the corrected Akaike Information Criterion (AIC$_\mathrm{c}$) and the Bayesian Information Criterion (BIC), as shown in Table. (\ref{tab:AICBIC_LCDMref}), using two combinations of datasets: Dataset~I and Dataset~II. In this work, we adopt the $\Lambda$CDM model as the reference to calculate the differences $\Delta\mathrm{AIC}_\mathrm{c}$ and $\Delta\mathrm{BIC}$. Regarding Dataset~I, we find that Model~I yields the lowest values for both AIC$_\mathrm{c}$ and BIC, with $\Delta\mathrm{AIC}_\mathrm{c} = -8.85$ and $\Delta\mathrm{BIC} = -3.40$, indicating that Model~I fits Dataset~I significantly better than Model~II and $\Lambda$CDM. A similar conclusion holds for Dataset~II, where Model~I remains the preferred model, with $\Delta\mathrm{AIC}_\mathrm{c} = -12.29$ and $\Delta\mathrm{BIC} = -6.91$.\\
\begin{figure}[h!]
    \centering
    \begin{minipage}[b]{0.44\textwidth}
        \centering
        \includegraphics[width=\textwidth]{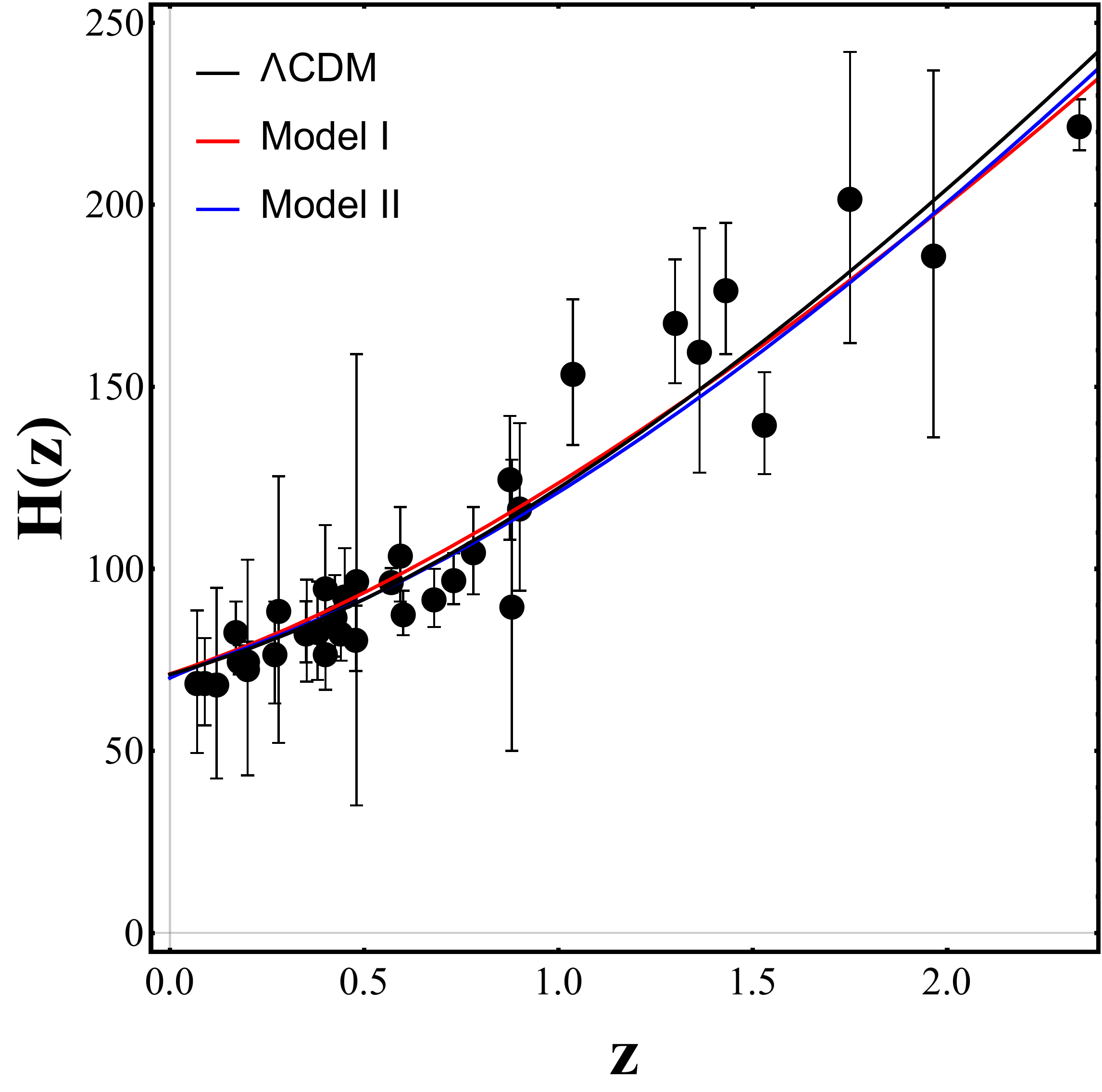}
        \caption{Error bar plots of 36 data points from the Cosmic Chronometers dataset are presented, along with the corresponding fits of the Hubble parameter $H(z)$ as a function of redshift $z$ for the $\Lambda$CDM model, Model~I, and Model~II.}
        \label{fig:cc}
    \end{minipage}
  \hspace{0.07\textwidth} 
    \begin{minipage}[b]{0.44\textwidth}
        \centering
        \includegraphics[width=\textwidth]{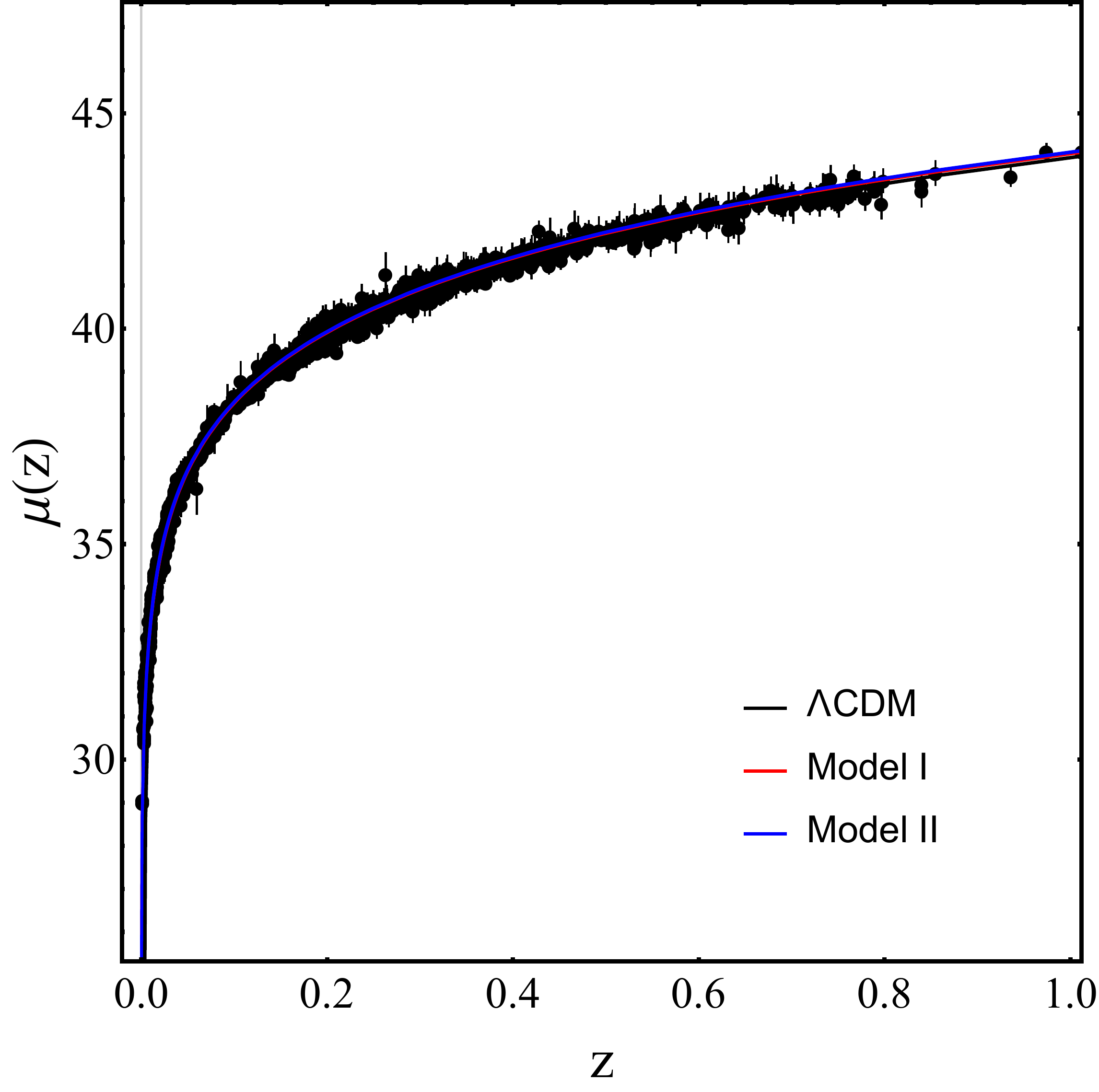}
        \caption{Error bar plots of 1701 data points from the Pantheon plus dataset are shown, displaying the fit of the distance modulus function $\mu(z)$ as a function of redshift $z$ for the $\Lambda$CDM model, Model~I, and Model~II.}
        \label{fig:sn}
    \end{minipage}
\end{figure}
\begin{table}[t!]
\centering
\renewcommand{\arraystretch}{1.2}
\begin{tabular}{lccc}
\hline
\textbf{} & \textbf{$\Lambda$CDM} & \textbf{$f(G)$ Model I} & \textbf{$f(G)$ Model II} \\
\hline
\multicolumn{4}{c}{\textbf{Dataset I}} \\
$AIC_{c}$         & 1570.14 & 1561.29 & 1563.69 \\
$\Delta AIC_{c}$ & 0       & $-8.85$ & $-6.45$ \\
$BIC $        & 1586.51 & 1583.11 & 1585.51 \\
$\Delta$BIC & 0       & $-3.40$ & $-1.00$ \\
\hline
\multicolumn{4}{c}{\textbf{Dataset II}} \\
$AIC_c$         & 1592.06 & 1579.77 & 1581.34 \\
$\Delta AIC_{c}$ & 0       & $-12.29$ & $-10.72$ \\
$BIC$         & 1613.91 & 1607 & 1608.64 \\
$\Delta$BIC & 0       & $-6.91$ & $-5.27$ \\
\hline
\end{tabular}
\caption{Model comparison using $AIC_{c}$ and $BIC$ for the $\Lambda$CDM and two $f(G)$ gravity models, based on the Dataset I and Dataset II. Differences ($\Delta AIC_{c}$, $\Delta BIC$) are shown relative to the $\Lambda$CDM model.}
\label{tab:AICBIC_LCDMref}
\end{table}
\begin{figure}[h!]
    \centering
    \includegraphics[width=0.6\textwidth]{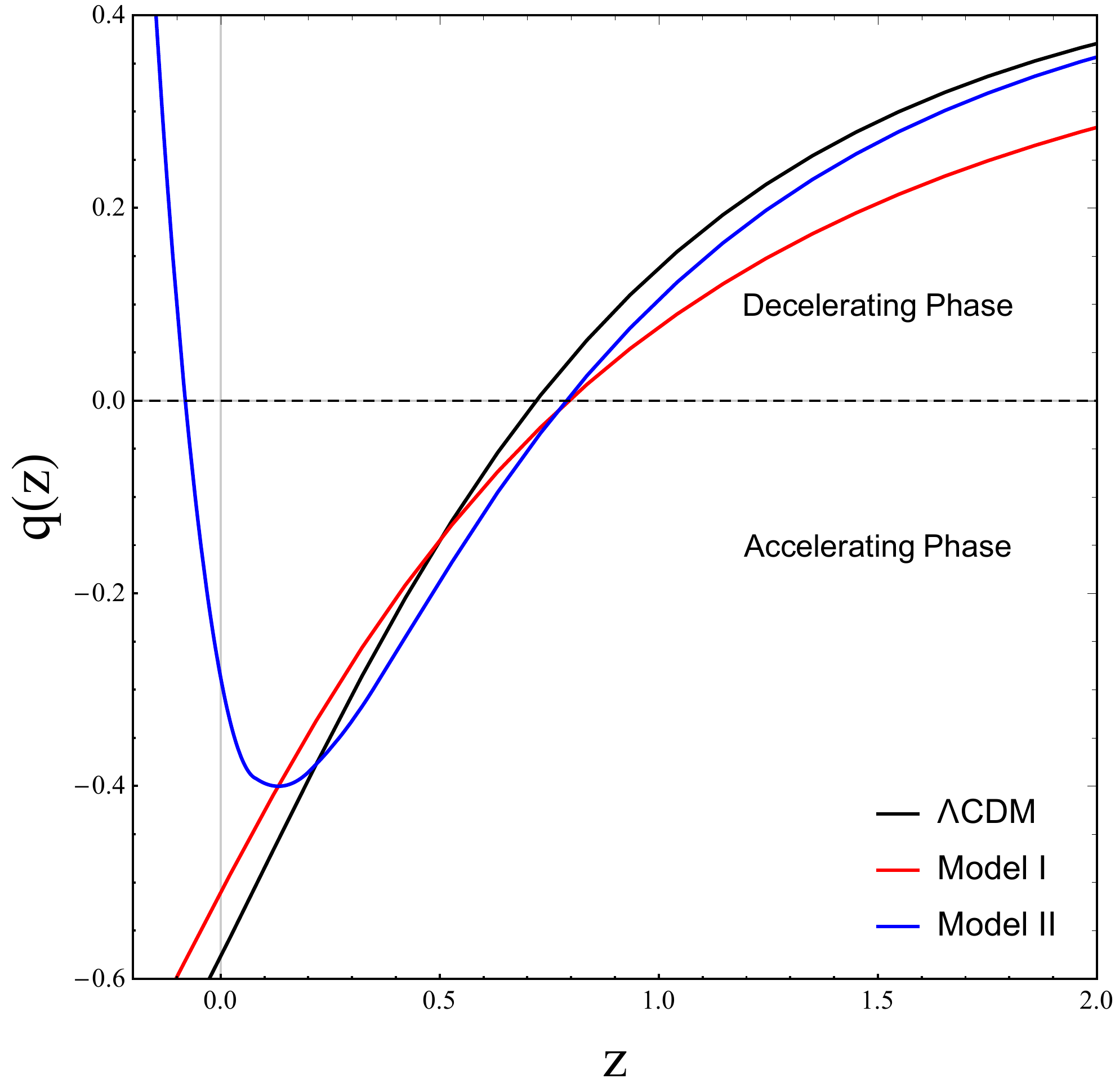}
    \caption{Graphical representation of deceleration parameter (Eq. (\ref{F33})) versus redshift using Dataset II (CC+PP+DESI BAO). The red line represents the behavior of deceleration parameter for the power law $f(G)$ model,  blue line represents the behavior of deceleration parameter for the exponential $f(G)$ model, while black line represents the behavior of deceleration parameter for the $\Lambda$CDM model.}
    \label{dec}
\end{figure}

Using the mean parameter values obtained in Table~\ref{Mean}, we plot the Hubble function $H(z)$ and the distance modulus $\mu(z)$ for the $\Lambda$CDM model and the two $f(G)$ models in Figures~\ref{fig:cc} and~\ref{fig:sn}, along with the corresponding error bars from cosmic chronometers and Pantheon plus data. We find that all three models are compatible with current observational data. In addition, Figure~\ref{dec} shows the evolution of the deceleration parameter $q(z)$ as a function of redshift for the $\Lambda$CDM model (black), Model I (red), and Model II (blue), using the combined Dataset II. The deceleration parameter $q(z)$ quantifies the rate at which the expansion of the Universe is accelerating or decelerating. Specifically, a positive $q(z)$ indicates decelerated expansion, while a negative $q(z)$ corresponds to accelerated expansion. Figure~\ref{dec} reveals that the Universe undergoes a transition from a decelerated to an accelerated expansion phase in the recent past, characterized by the redshift at which $q(z_{\text{t}}) = 0$. The transition redshift occurs at $z_t = 0.793$ for Model I, $z_t = 0.788$ for Model II, and $z_t = 0.720$ for the $\Lambda$CDM model. These transition redshifts are in good agreement with existing observational constraints, as reported in \cite{lohakare2024cosmology,gruber2014cosmographic,capozziello2014cosmographic,farooq2013hubble,munyeshyaka2023perturbations}. We also obtain the present-day values of the deceleration parameter as $q_0 = -0.51$ for Model I, $q_0 = -0.287$ for Model II, and $q_0 = -0.577$ for the $\Lambda$CDM model. These values are consistent with recent observational estimates, particularly the range $q_0 = -0.528^{+0.092}_{-0.088}$ reported in \cite{lohakare2024cosmology,gruber2014cosmographic,capozziello2014cosmographic}. A noteworthy result in Figure~\ref{dec} is that, for Model II, the Universe is predicted to undergo multiple phases of transition specifically, a future transition from accelerated to decelerated expansion at a redshift of approximately $z \approx -0.1$, in contrast to Model I and the $\Lambda$CDM model. This prediction is consistent with recent findings reported in \cite{enkhili2024cosmological,futur}.

\section{Conclusions}\label{con}
In the present work, we have observationally investigated the impact of $f(G)$ gravity models on the cosmological expansion history of the Universe, focusing on two specific functional forms of $f(G)$. The first is a power-law model given by $f(G) = \alpha G^{\beta}$ (Model~I), and the second is an exponential model given by $f(G) = \alpha G_0\left(1 - e^{-p \frac{G}{G_0}}\right)$ (Model~II), where $\alpha$, $\beta$, and $p$ are free model parameters. These two forms are constructed such that the standard $\Lambda$CDM model is recovered in the limits $\beta = 0$ for Model~I and $p \to \infty$ for Model~II, as can be verified via Eqs.~(\ref{F11}) and~(\ref{F22}), respectively. After deriving and numerically solving the normalized Friedmann equations for both models, we performed a Markov Chain Monte Carlo (MCMC) analysis to constrain the model parameters using two combinations of observational datasets: Dataset~I (PP+CC) and Dataset~II (PP+CC+DESI BAO). For Dataset~I, Model~I yields the mean parameter values: $\Omega_{m} = 0.190^{+0.041}_{-0.036}$, $h = 0.7149 \pm 0.0086$, and $\beta = 0.260^{+0.079}_{-0.069}$. For Model~II, we obtain $\Omega_{m} = 0.262 \pm 0.014$, $h = 0.711 \pm 0.0087$, and $p = 3.97^{+0.38}_{-0.47}$. For Dataset~II, Model~I gives: $\Omega_{m} = 0.230^{+0.027}_{-0.024}$, $h = 0.7115 \pm 0.0083$, and $\beta = 0.192^{+0.068}_{-0.062}$, while Model~II yields: $\Omega_{m} = 0.275 \pm 0.011$, $h = 0.7085 \pm 0.0084$, and $p = 4.18^{+0.35}_{-0.40}$. For comparison, the $\Lambda$CDM model used as a reference yields $\Omega_{m} = 0.282 \pm 0.0129$, $h = 0.7141 \pm 0.0087$ using Dataset~I, and $\Omega_{m} = 0.280 \pm 0.011$, $h = 0.7139 \pm 0.0082$ using Dataset~II. A summary of the best-fit values, associated uncertainties, and key results for the free parameters $\beta$ and $p$, as well as for $\Omega_{m}$ and the dimensionless Hubble parameter $h$, for both $f(G)$ gravity models using Dataset~I and Dataset~II, is presented in Table~\ref{Mean}.\\

Using the MCMC analysis, we obtain the confidence contours for each dataset combination. Figure~\ref{ca} shows the 1$\sigma$ and 2$\sigma$ contour plots for Model~I of the $f(G)$ gravity model, illustrating the constraints on the parameters $\beta$, $\Omega_{m}$, and $h$ from Datasets~I and~II. For Model~II, the corresponding contour plots for the parameters $p$, $\Omega_{m}$, and $h$ are displayed in Fig.~\ref{c2}. In Fig.~\ref{c3}, we present a comparative analysis of the contours for Model~I, Model~II, and the $\Lambda$CDM model based on Dataset~II. A negative correlation is observed between $\Omega_{m}$ and $\beta$, whereas a positive correlation appears between $\Omega_{m}$ and $p$. The next task was to perform a statistical analysis incorporating the corrected Akaike Information Criterion (AIC$_{\rm c}$), the Bayesian Information Criterion (BIC), as well as their relative differences $\Delta$AIC$_{\rm c}$ and $\Delta$BIC. Using Dataset~I, we find that Model~I yields $\Delta$AIC$_{\rm c} = -8.85$ and $\Delta$BIC $= -3.40$, while for Model~II the values are $\Delta$AIC$_{\rm c} = -6.45$ and $\Delta$BIC $= -1.00$. For the $\Lambda$CDM model, which is used as the reference, both $\Delta$AIC$_{\rm c}$ and $\Delta$BIC are by definition equal to zero. Considering Dataset~II, Model~I provides $\Delta$AIC$_{\rm c} = -12.29$ and $\Delta$BIC $= -6.91$, whereas Model~II gives $\Delta$AIC$_{\rm c} = -10.72$ and $\Delta$BIC $= -5.27$. A summary of this comparison is presented in Table~\ref{tab:AICBIC_LCDMref}. As shown in the table, both $\Delta$AIC$_{\rm c}$ and $\Delta$BIC values are negative for Model~I and Model~II across both datasets, this suggests that, within the considered data sets, the $f(G)$ models are statistically favored over $\Lambda$CDM. This results was also obtained using DESI BAO in Refs. \cite{Dynamical} and \cite{Modified} for $f(R)$ gravity and in Ref. \cite{Einstein} for the Einstein-Gauss-Bonnet cosmology. Moreover, the comparison reveals that Model~I is preferred over the Model~II for both datasets, based on both $\Delta$AIC$_{\rm c}$ and $\Delta$BIC criteria. These results are in line with findings reported in~\cite{mhamdi2024cosmological,briffa2023f} for $f(Q)$ and $f(T)$ gravity models, respectively. Similarly, studies such as~\cite{sahlu2025structure,anagnostopoulos2021first} also favor power-law models over exponential ones, while other works like~\cite{capozziello2022model,mhamdi2024constraints} suggest the opposite preference. Overall, our results provide strong evidence that the $f(G)$ gravity framework can serve as a viable alternative to $\Lambda$CDM in describing the dynamics of the universe without invoking the dark energy hypothesis.\\

Using the obtained mean values of the model parameters, we analyzed the dynamics of the universe by plotting the deceleration parameter $q(z)$, as shown in Fig.~\ref{dec}. The present-day values of the deceleration parameter are found to be $q_0 = -0.51$ for Model~I, $q_0 = -0.287$ for Model~II, and $q_0 = -0.577$ for the $\Lambda$CDM model. The transition redshift, defined as the redshift at which the universe transitions from deceleration to acceleration, is determined to be $z_t = 0.793$ for Model~I, $z_t = 0.788$ for Model~II, and $z_t = 0.720$ for the $\Lambda$CDM model. It is worth noting that Model~II exhibits an additional transition in the future (at $z \approx -0.1$), indicating a possible return to a decelerating phase. This behavior contrasts with both Model~I and the $\Lambda$CDM model, which continue to exhibit accelerated expansion into the future.\\

To further test the viability of the proposed models in $f(G)$ gravity, we plan to extend our analysis by incorporating additional observational data, such as Redshift Space Distortion (RSD) measurements and the Union3 supernova compilation. We also aim to explore the perturbative regime of these models to study the growth of large-scale structures using RSD data. In addition, we intend to examine the implications of $f(G)$ gravity for current cosmological tensions, including the $H_0$ and $S_8$ discrepancies, by incorporating Cosmic Microwave Background (CMB) data. These extensions will be pursued in future work.
\acknowledgments
PKD wish to mention that the part of the numerical computation of this work was carried out on the Pegasus Computing Cluster of IUCAA, Pune, India and also acknowledges the Inter-University Centre for Astronomy and Astrophysics (IUCAA), Pune, India, for providing him a Visiting Associateship under which a part of this work was carried out. AM acknowledges the hospitality of the University of Rwanda-College of Science and Technology, where part of this work was conceptualised and completed. JN thanks Rwanda Astrophysics, Space and Climate Science Research Group for the support.

\end{document}